\documentclass[preprintnumbers,amsmath,amssymbm,preprint]{revtex4}
\usepackage{epsfig}
\usepackage{graphicx}

\begin{document}
\title{Purely imaginary polar resonances of rapidly-rotating Kerr black holes}
\author{Shahar Hod}
\address{The Ruppin Academic Center, Emeq Hefer 40250, Israel}
\address{ }
\address{The Hadassah Institute, Jerusalem 91010, Israel}
\date{\today}

\begin{abstract}
\ \ \ We prove the existence of a unique family of non-oscillatory
(purely-imaginary) polar quasinormal resonances of rapidly-rotating
Kerr black holes. These purely imaginary resonances can be expressed
in the compact form: $\omega_n=-i2\pi T_{\text{BH}}(l+1+n)$, where
$T_{\text{BH}}$ is the black-hole temperature, $l$ is the spheroidal
harmonic index of the mode, and $n=0,1,2,...$ is the resonance
parameter. It is shown that our {\it analytical} results for the
black-hole resonance spectrum agree with new {\it numerical} data
that recently appeared in the literature.
\end{abstract}
\bigskip
\maketitle

%]

\section{Introduction}

Astrophysically realistic black holes are expected to possess a
non-zero spin angular momentum. In fact, recent astrophysical
observations \cite{Bervol,Bre,Rey} suggest that rapidly-rotating
black holes are ubiquitous in our Universe \cite{Yang3}. It is
therefore of astrophysical importance to explore the physical
properties of rapidly-spinning black holes.

Rapidly-rotating black holes are also important from the point of
view of quantum field theory. In particular, the conjectured
relation between the near-horizon quantum states of a near-extremal
black hole and the quantum states of a two-dimensional conformal
field theory enables one to count the entropy of these black hole
\cite{Bar,Gui,Stro}. Moreover, near-extremal black holes have the
interesting property of saturating the recently conjectured
universal relaxation bound \cite{Hod1b,Gruzb,Pescib,Hodbb1,Hodbb2}.

In the present study we shall explore the resonance spectrum of
rapidly-rotating (near-extremal) black holes. Quasinormal (QNM)
resonances are the unique `sound' of black holes. (See
\cite{Nollert1,Ber1,Kono} for excellent reviews and detailed lists
of references.) Perturbed black holes usually display a
characteristic pattern of damped oscillations of the form
$e^{-i\omega_{\text{R}} t-\omega_{\text{I}}t}$. These damped
oscillations characterize the decay (relaxation) of the perturbation
fields in the black-hole spacetime (note that a decaying
perturbation is characterized by a complex frequency
$\omega=\omega_{\text{R}}-i\omega_{\text{I}}$ with
$\omega_{\text{I}}>0$). The spectrum of quasinormal resonances
reflects the physical properties (mass and angular momentum) of the
black hole itself \cite{Notetail,Tails1}.

The quasinormal resonances correspond to wave fields propagating in
the black-hole spacetime with the physically motivated boundary
conditions of purely ingoing waves crossing the black-hole horizon
and purely outgoing waves at spatial infinity \cite{Detw}. For given
values of the angular parameters $m$ and $l$ [here $m$ is the
azimuthal harmonic index and $l$ is the spheroidal harmonic index of
the perturbation mode, see Eq. (\ref{Eq5}) below] these boundary
conditions single out a countable set
$\{\omega(n;m,l)\}_{n=0}^{n=\infty}$ of complex black-hole
resonances. The fundamental quasinormal resonance, the mode with the
smallest imaginary part, determines the characteristic timescale
$\tau_{\text{relax}}$ for generic black-hole perturbations to decay:
$\tau_{\text{relax}}\sim 1/\omega_{\text{I}}(n=0)$
\cite{Hod1b,Hodbb1,Hodbb2}.

The characteristic spectrum of black-hole resonances is of
fundamental importance from both the astrophysical
\cite{Nollert1,Ber1,Kono} and theoretical
\cite{HodPRL,Gary,Hodqg,KeshHod,KeshNe,Hart} points of view. In
particular, these characteristic oscillations are expected to be
excited in a variety of astrophysical processes involving black
holes \cite{Nollert1,Ber1,Kono}. The excitation of black-hole
quasinormal resonances thus provides a promising observational way
for identifying the physical parameters (masses and angular momenta)
of astrophysical black holes.

It is worth noting that the characteristic resonance spectra of most
black-hole spacetimes must be computed numerically
\cite{Nollert1,Ber1,Kono,Noll2}. [See, however,
\cite{Mash,Goeb,CarMir,Dolan,Massfr,Yang,Hodpl} for some interesting
analytical results in the asymptotic limit $l\gg1$ (the
geometric-optics approximation with $\omega_{\text{R}}\gg1$), and
\cite{HodPRL,Hodqg,KeshHod,KeshNe} for interesting analytical
results in the highly-damped asymptotic regime $n\gg1$ (the
$\omega_{\text{I}}\gg1$ limit)]. A notable exception is provided by
the family of  rapidly-rotating (near-extremal) black holes with
$a\simeq M$. (Here $M$ and $Ma$ are the mass and angular momentum of
the black holes, respectively). In particular, it was shown in
\cite{Hod1b,Hodbb2} that, for rapidly-rotating black holes, the
resonance spectrum of {\it co-rotating} ($m>0$) modes is described
by the remarkably simple analytic formula \cite{Noteunit}:
\begin{equation}\label{Eq1}
\omega(n;l,m)=m\Omega-i2\pi T_{\text{BH}}(n+{1\over 2}-i\delta)\  ,
\end{equation}
where
\begin{equation}\label{Eq2}
T_{\text{BH}}\equiv {{r_+-r_-}\over{4\pi(r^2_++a^2)}}\ \ \
\text{and} \ \ \ \Omega\equiv {{a}\over{r^2_++a^2}}
\end{equation}
are the physical parameters of the black hole: its temperature and
angular velocity, respectively. [Here $r_{\pm}\equiv
M\pm\sqrt{M^2-a^2}$ are the radii of the black-hole horizons]. The
parameter $\delta=\delta(l,m)$ is closely related to the angular
eigenvalue of the corresponding angular equation, see Eq.
(\ref{Eq5}) below.

It is worth emphasizing that the family (\ref{Eq1}) of co-rotating
$m>0$ resonances describes perturbations of the rapidly-rotating
(near-extremal) black-hole spacetimes with extremely long relaxation
times, $\tau_{\text{relax}}\sim 1/\omega_I=O(T^{-1}_{\text{BH}})\gg
M$.

\section{Recently published numerical results}

Recently, Yang et. al. \cite{Yang3} have computed numerically the
{\it polar} ($m=0$) quasinormal resonances of near-extremal Kerr
black holes. Remarkably, the numerical data presented in
\cite{Yang3} (see, in particular, Fig. 7 of \cite{Yang3}) reveals
the existence of a previously unknown family of polar resonances
which are quite accurately described by the same formula (\ref{Eq1})
with $m=0$, namely \cite{Notedel}:
\begin{equation}\label{Eq3}
\omega^{\text{numerical}}(n;m=0)\simeq -i2\pi T_{\text{BH}}(n+l+1)\
.
\end{equation}

It should be noted that the numerical results of Yang. et. al.
\cite{Yang3} regarding the polar black-hole resonance spectrum
(\ref{Eq3}) are quite surprising for two reasons:
\newline
(1) First of all, it should be emphasized that former analytical
techniques and approximations which were used in
\cite{Hod1b,Hodbb2,TeuPre1,TeuPre2,Star,Det} in order to derive the
analytical formula (\ref{Eq1}) for the co-rotating $m>0$ resonances
[which are characterized by $M\omega=O(1)$] are {\it no} longer
valid (and therefore cannot be applied) in the low frequency
$M\omega\ll1$ regime of the polar $m=0$ resonances (\ref{Eq3}), see
Appendix A for details.
\newline
(2) To the best of our knowledge, all former numerical studies of
the Kerr quasinormal spectrum (see ,in particular, the numerical
studies of the Kerr spectrum in Refs. \cite{Leaver,Ono1,Ander,Ono2})
have not reported on the existence of these non-oscillatory (purely
imaginary, $\omega_{\text{R}}=0$) polar resonances. In particular,
former numerical studies \cite{Leaver,Ono1,Ander,Ono2} have claimed
that the fundamental polar resonances are characterized by finite
real ($\omega_{\text{R}}\neq0$) oscillation frequencies. It thus
seems that there is some controversy in the literature regarding the
nature of the black-hole polar resonances.

It is therefore of physical importance to prove (or disprove)
analytically the existence of this unique family of non-oscillatory
(purely imaginary) polar quasinormal resonances. The main goal of
the present work is to provide such an {\it analytical} proof. In
particular, as we shall show below, the spectrum of polar black-hole
resonances can be studied analytically in the double limit
$M\omega\ll1$ with $MT_{\text{BH}}\ll1$.

\section{Description of the system}

The physical system we consider consists of a massless spin-s
\cite{Notespin} field linearly coupled to a Kerr black hole of mass
$M$ and angular momentum per unit mass $a$. In order to facilitate a
fully analytical study, we shall assume that the black hole is
rapidly-rotating with $a\simeq M$.

Teukolsky \cite{Teu} has shown that the dynamics of a massless
spin-s field $\Psi$ in the rotating Kerr black-hole spacetime can be
described by a single master equation. Decomposing the field $\Psi$
in the form
\begin{equation}\label{Eq4}
{_s\Psi_{lm}}(t,r,\theta,\phi)=e^{im\phi}{_sS_{lm}}(\theta;a\omega){_s\psi_{lm}}(r)e^{-i\omega
t}\ ,
\end{equation}
one finds \cite{Teu} that the functions ${_s\psi_{lm}}$ and
${_sS_{lm}}$ obey radial and angular equations of the confluent Heun
type \cite{Teu,Heun,Flam,Fiz1} which are coupled by a separation
constant $_sA_{lm}(a\omega)$. Here $(t,r,\theta,\phi)$ are the
Boyer-Lindquist coordinates, $\omega$ is the (conserved) frequency
of the mode, $m$ is the azimuthal harmonic index of the mode, and
$l$ is the spheroidal harmonic index with
$l\geq\text{max}\{|m|,|s|\}$. We shall henceforth study the polar
$m=0$ sector of black-hole perturbations.

The angular functions ${_sS_{l0}}(\theta;a\omega)$ are known as the
spin-weighted spheroidal harmonics. These function satisfy the
angular equation \cite{Teu,Heun,Flam}
\begin{equation}\label{Eq5}
{1\over {\sin\theta}}{\partial \over
{\partial\theta}}\Big(\sin\theta {{\partial
S}\over{\partial\theta}}\Big)+\Big(a^2\omega^2\cos^2\theta-2a\omega
s\cos\theta-s^2\cot^2\theta+s+{_sA_{l0}}-a^2\omega^2\Big)S=0\
\end{equation}
and are required to be regular at the poles $\theta=0$ and
$\theta=\pi$. These boundary conditions pick out a discrete set of
eigenvalues $\{_sA_{l0}\}$ labeled by the integer $l$.

In the small frequency $a\omega\ll 1$ regime, the regime we are
interested in in this study, the angular functions can be
approximated by the familiar spin-weighted spherical harmonics. The
angular eigenvalues can then be expanded in the form
\cite{TeuPre1,BerAlm}
\begin{equation}\label{Eq6}
_sA_{l0}=l(l+1)-s(s+1)+\epsilon(a\omega)^2+O(a^4\omega^4)\ ,
\end{equation}
where the expansion coefficient $\epsilon=\epsilon(l,s)$ is given by
\cite{TeuPre1,BerAlm}
\begin{equation}\label{Eq7}
\epsilon(l,s)=-2{{3s^4-2s^2l(l+1)-l(l+1)(l^2+l-1)}\over{l(l+1)(2l-1)(2l+3)}}\
.
\end{equation}
Thus, in the $a\omega\ll1$ limit one can write
\begin{equation}\label{Eq8}
{_sA_{l0}}=\ell(\ell+1)-s(s+1)\  ,
\end{equation}
where
\begin{equation}\label{Eq9}
\ell\equiv l+{{\epsilon}\over{2l+1}}(a\omega)^2+O(a^4\omega^4)
\end{equation}
is nearly an integer with a small correction term of order
$O(a^2\omega^2)$.
%\begin{equation}\label{Eq7}
%\epsilon(l,s)=-{{6s^4-4s^2l(l+1)+l(l+1)(2l^2+2l-1)}\over{l(l+1)(2l-1)(2l+3)}}\
%.
%\end{equation}
(We shall henceforth omit the indices $s$ and $l$ for brevity.)

The radial function $\psi(r)$ satisfies the differential equation
\cite{Teu}
\begin{equation}\label{Eq10}
\Delta^{-s}{{d}
\over{dr}}\Big(\Delta^{s+1}{{d\psi}\over{dr}}\Big)+\Big[{{(r^2+a^2)^2\omega^2-2is\omega
M(r^2-a^2)}\over{\Delta}}+2is\omega r-A\Big]\psi=0\ ,
\end{equation}
where $\Delta\equiv r^2-2Mr+a^2$. The scattering process of massless
fields in the black-hole spacetime is governed by the radial wave
equation (\ref{Eq10}) supplemented by the physically motivated
boundary conditions of purely ingoing waves crossing the black-hole
horizon and a mixture of both ingoing (incident) and outgoing
(reflected) waves at spatial infinity \cite{Nollert1,Ber1,Kono}:
\begin{equation}\label{Eq11}
%\label{eq:boundary_conditions}
\psi \sim
\begin{cases}
e^{-i\omega y}+{\mathcal{R}}(\omega)e^{i \omega y} & \text{ as }
r\rightarrow\infty\ \ (y\rightarrow \infty)\ ; \\
{\mathcal{T}}(\omega)e^{-i\omega y} & \text{ as } r\rightarrow r_+\
\ (y\rightarrow -\infty)\ ,
\end{cases}
\end{equation}
where the ``tortoise" radial coordinate $y$ is defined by
$dy=[(r^2+a^2)/\Delta]dr$. The frequency-dependent functions ${\cal
R}(\omega)$ and ${\cal T}(\omega)$ represent the reflection and
transmission amplitudes for a field of conserved frequency $\omega$
coming from infinity.
%These scattering amplitudes satisfy the probability conservation
%equation $|{\cal T}(\omega)|^2+|{\cal R}(\omega)|^2=1$.

\section{Analytic derivation of the quasinormal resonances}

The spectrum of black-hole quasinormal resonances is associated with
the poles of the (frequency dependent) transmission and reflection
amplitudes \cite{Detw,Leaver}. As we shall now show, these
characteristic resonances can be studied analytically in the double
limit
\begin{equation}\label{Eq12}
M\omega\ll 1 \ \ \ \text{and} \ \ \ MT_{\text{BH}}\ll1
\end{equation}
of small frequencies and small black-hole temperatures (that is,
rapidly-spinning black holes with $a\simeq M$).

To that end, we shall follow the analysis of \cite{Chan,Page,Hodcen}
in order to calculate the scattering amplitudes in the low frequency
regime $M\omega\ll1$. It is convenient to define new dimensionless
variables
\begin{equation}\label{Eq13}
x\equiv {{r-r_+}\over {r_+-r_-}}\ \ ;\ \
\varpi\equiv{{\omega}\over{2\pi T_{BH}}}\ \ ;\ \ k\equiv
\omega(r_+-r_-)\  ,
\end{equation}
in terms of which the radial wave equation (\ref{Eq10}) becomes
\begin{eqnarray}\label{Eq14}
x^2(x+1)^2{{d^2\psi}\over{dx^2}}+(s+1)x(x+1)(2x+1){{d\psi}\over{dx}}\\
\nonumber
+\big[k^2x^4+2iskx^3-Ax(x+1)-is\varpi(x+1/2)+\varpi^2/4\big]\psi&=&0\
.
\end{eqnarray}

The solution of the radial wave equation (\ref{Eq14}) in the
near-horizon regime $kx\ll 1$ which satisfies the physical boundary
condition of purely ingoing waves crossing the black-hole horizon is
given by \cite{Chan,Page,Hodcen}
\begin{eqnarray}\label{Eq15}
\psi=x^{-s-i\varpi/2}(x+1)^{-s+i\varpi/2}
{_2F_1}(-\ell-s,\ell-s+1;1-s-i\varpi;-x) \ ,
\end{eqnarray}
where $_2F_1(a,b;c;z)$ is the hypergeometric function \cite{Abram}.

The solution of the radial wave equation (\ref{Eq14}) in the
far-region $x\gg |\varpi| +1$ is given by \cite{Chan,Page,Hodcen}
\begin{eqnarray}\label{Eq16}
\psi=Ae^{-ikx}x^{\ell-s}{_1F_1}(\ell-s+1;2\ell+2;2ikx)+Be^{-ikx}x^{-\ell-s-1}{_1F_1}(-\ell-s;-2\ell;2ikx)\
,
\end{eqnarray}
where $_1F_1(a;c;z)$ is the confluent hypergeometric function
\cite{Abram}.

The coefficients $A$ and $B$ can be determined by matching the
near-horizon solution (\ref{Eq15}) with the far-region solution
(\ref{Eq16}) in the overlap region $|\varpi|+1\ll x\ll 1/k$. This
matching procedure yields \cite{Chan,Page,Hodcen}
\begin{equation}\label{Eq17}
A={{\Gamma(2\ell+1)\Gamma(1-s-i\varpi)}\over
{\Gamma(\ell-s+1)\Gamma(\ell+1-i\varpi)}}\
 ,
\end{equation}
and
\begin{equation}\label{Eq18}
B={{\Gamma(-2\ell-1)\Gamma(1-s-i\varpi)}\over
{\Gamma(-\ell-s)\Gamma(-\ell-i\varpi)}}\ .
\end{equation}
Finally, the asymptotic ($x\gg1$) form of the confluent
hypergeometric functions \cite{Abram} can be used in order to write
the far-region solution (\ref{Eq16}) in the form given by Eq.
({\ref{Eq11}). After some algebra one finds \cite{Chan,Page,Hodcen}
\begin{equation}\label{Eq19}
|{\cal T}(\omega)|^2=\Re \Big\{4e^{i\pi(s+1/2)}\cos[(\ell-s)\pi]
\Big[{{\Gamma(-2\ell-1)\Gamma(\ell-s+1)}\over{\Gamma(2\ell+1)\Gamma(-\ell-s)}}\Big]^2
{{\Gamma(\ell+1-i\varpi)}\over{\Gamma(-\ell-i\varpi)}}(2k)^{2\ell+1}\Big\}\
,
\end{equation}
for the transmission probability.

The quasinormal modes represent the scattering resonances of the
fields in the black-hole spacetime. These characteristic resonances
correspond to the poles of the (frequency dependent) transmission
and reflection amplitudes \cite{Detw,Leaver}. The well-known pole
structure of the Gamma functions \cite{Abram} implies that the
transmission probability (\ref{Eq19}) has poles at
\begin{equation}\label{Eq20}
\ell+1-i\varpi=-n\  ,
\end{equation}
where $n=0,1,2,...$ is the resonance parameter. Equation
(\ref{Eq20}) represents the resonance condition for the
characteristic quasinormal frequencies of the black-hole spacetime.

Taking cognizance of Eq. (\ref{Eq13}), one can express the polar
black-hole resonances in the simple form
\begin{equation}\label{Eq21}
\omega_n=-i2\pi T_{\text{BH}}(\ell+1+n)\  .
\end{equation}
It is worth emphasizing again that the resonance spectrum
(\ref{Eq21}) is valid in the small frequency regime $M\omega\ll1$.
This implies that the spectrum (\ref{Eq21}) is valid in the regime
of rapidly-rotating (near-extremal) black holes with
$MT_{\text{BH}}\ll1$ \cite{Noterel}. Note also that the analytically
derived polar spectrum (\ref{Eq21}) agrees with the recently
published numerical data of Yang. et. al. \cite{Yang3}.

\section{Summary}

Motivated by the recently published numerical results of Yang et.
al. \cite{Yang3}, we have studied analytically the polar ($m=0$)
resonance spectrum of rapidly-rotating (near-extremal) black holes.
The numerical results reported by Yang et. al. \cite{Yang3} are
quite surprising: while former numerical studies of the Kerr
quasinormal spectrum \cite{Leaver,Ono1,Ander,Ono2} have reported on
polar quasinormal resonances which are characterized by finite
oscillations frequencies ($\omega_{\text{R}}\neq0$), the new
numerical study of \cite{Yang3} has revealed the existence of a
different branch of purely imaginary (non-oscillatory,
$\omega_{\text{R}}=0$) polar quasinormal resonances.

This new branch of polar quasinormal resonances seems to have been
overlooked in former numerical studies of the Kerr quasinormal
spectrum \cite{Leaver,Ono1,Ander,Ono2}. Moreover, we have shown (see
Appendix A) that former analytical techniques and approximations
which were used in \cite{Hod1b,Hodbb2,TeuPre2,Det} in order to study
the co-rotating $m>0$ resonances of rapidly-rotating black holes are
{\it not} valid (and therefore cannot be applied) in the low
frequency regime of the polar $m=0$ quasinormal resonances.

Our main goal in the present study was to prove {\it analytically}
the existence of this unique family of non-oscillatory
(purely-imaginary) black-hole resonances. Using an appropriate small
frequency $M\omega\ll1$ approximation for the Teukolsky wave
equation (instead of the $\omega-m\Omega\ll\omega$ approximation
which was used in \cite{Hod1b,Hodbb2,TeuPre2,Det} in order to study
the co-rotating $m>0$ resonances), we have established the existence
of the new family (\ref{Eq21}) of purely-imaginary polar $m=0$
quasinormal frequencies.

Finally, it is interesting to analyze the complex spectrum of total
reflection modes (TRM) which characterizes the black-hole spacetime.
These important frequencies are determined by the condition ${\cal
T}(\omega^{\text{TRM}})=0$. Taking cognizance of Eq. (\ref{Eq19}),
one finds that the TRMs are determined by the requirement
$1/\Gamma(-\ell-i\varpi)=0$ [see Eq. (\ref{Eq19}) with ${\cal
T}(\omega)=0$]. This condition yields the simple spectrum
\begin{equation}\label{Eq22}
\omega^{\text{TRM}}_n=-i2\pi T_{\text{BH}}(-\ell+n)\
\end{equation}
for the TRMs of the black-hole spacetime.

Taking cognizance of Eqs. (\ref{Eq21}) and (\ref{Eq22}), one
realizes that the two spectra are almost identical: there are close
pairs of QNMs and TRMs. In particular, the difference between a QNM
frequency of overtone index $n$ and a nearby TRM frequency of
overtone index $n'=n+2l+1$ is extremely small, of the order of
$O(T^3_{\text{BH}})$ \cite{Noteqt}. This observation implies that
one must use a numerical scheme of extreme precision in order to
compute the QNM frequencies numerically and, in particular, in order
to distinguish them numerically from the TRM frequencies. The close
proximity between the frequencies (\ref{Eq21}) of the quasinormal
resonances and the frequencies (\ref{Eq22}) of the total reflection
modes may explain the failure of previous numerical studies to
observe this unique family of purely imaginary polar resonances
\cite{NoteYang3}.

\bigskip
\noindent
{\bf ACKNOWLEDGMENTS}
\bigskip

This research is supported by the Carmel Science Foundation. I thank
Yael Oren, Arbel M. Ongo, and Ayelet B. Lata for helpful
discussions.

\bigskip

\newpage
\setcounter{equation}{0}
\renewcommand{\theequation}{A\arabic{equation}}

\begin{appendix}\label{App1}
{\bf Appendix A: The range of validity of former analytical studies}

In this Appendix we shall discuss the range of validity of the
resonance spectrum (\ref{Eq1}). This resonance spectrum, originally
derived in \cite{Hodbb2}, stems from the near-extremal resonance
condition of Detweiler \cite{Det}, see Eq. (9) of \cite{Det}. The
resonance condition (9) of \cite{Det} is based on an earlier
analysis of Teukolsky and Press, see Appendix A of \cite{TeuPre2}.
Thus, in order to find the range of validity of the resonance
condition (9) of \cite{Det}, one should carefully examine the range
of validity of the analysis presented in Appendix A of
\cite{TeuPre2}.

By carefully repeating the analysis of \cite{TeuPre2} step-by-step,
one realizes that the transition from Eq. (A3) of \cite{TeuPre2} to
Eq. (A4) of \cite{TeuPre2} (see Appendix A of \cite{TeuPre2}) is
based on the assumption
\begin{equation}\label{EqA1}
\omega x\gg \omega-m\Omega\  ,
\end{equation}
where
\begin{equation}\label{EqA2}
x\equiv (r-r_+)/r_+
\end{equation}
in the notations of \cite{TeuPre2}. [In particular, when moving from
Eq. (A3) to Eq. (A4), the authors of \cite{TeuPre2} keep the term
$4i\omega r_+x$ in the coefficient of $dR/dx$ but neglect the terms
$4iM(\omega-m\Omega)$ and $-(s+1)\sigma$ in this same coefficient,
where $\sigma\equiv (r_+-r_-)/r_+$ in the notations of
\cite{TeuPre2}. This approximation is valid provided $\omega x\gg
\text{max}(\omega-m\Omega,\sigma/M)$]. In addition, the transition
from Eq. (A3) of \cite{TeuPre2} to Eq. (A8) of \cite{TeuPre2} is
based on the explicit assumption
\begin{equation}\label{EqA3}
x\ll1\  .
\end{equation}

Thus, the dimensionless variable $x$ is required to satisfy the two
inequalities $(\omega-m\Omega)/\omega\ll x \ll 1$. These two
requirements can only be satisfied simultaneously in the regime
\begin{equation}\label{EqA4}
\omega-m\Omega\ll\omega\  .
\end{equation}
The inequality (\ref{EqA4}) is indeed satisfied by the co-rotating
$m>0$ modes (\ref{Eq1}) [these modes are characterized by
$M\omega=O(1)$ and $M(\omega-m\Omega)=O(MT_{\text{BH}})\ll1$]. On
the other hand, the inequality (\ref{EqA4}) is obviously {\it
violated} by the family (\ref{Eq3}) of polar $m=0$ quasinormal
modes.

One therefore concludes that the near-extremal resonance condition,
Eq. (9) of \cite{Det}, is {\it not} valid in the low frequency
regime of polar modes. Thus, in the present study we shall use a
different analytical approach in order to prove the existence of the
polar family (\ref{Eq3}) of quasinormal resonances. This alternative
analytical approach is based on a low frequency approximation
$M\omega\ll1$ to the Teukolsky wave equation (as opposed to the
$\omega-m\Omega\ll\omega$ approximation used in Appendix A of
\cite{TeuPre2}). This low frequency approximation is suitable for
the analysis of the polar resonance spectrum (\ref{Eq3}) in the near
extremal $MT_{\text{BH}}\ll1$ regime.

\end{appendix}
\maketitle


\bigskip

\begin{thebibliography}{99}

\bibitem{Bervol} E. Berti and M. Volonteri, Astrophys. J. {\bf 684}, 822 (2008).

\bibitem{Bre} L. Brenneman, C. Reynolds, M. Nowak, R. Reis, M. Trippe, et al.,
Astrophys.J. {\bf 736}, 103 (2011).

\bibitem{Rey} C. S. Reynolds, e-print arXiv:1307.3246.

\bibitem{Yang3} H. Yang, A. Zimmerman, A. Zenginoglu, F. Zhang, E. Berti, and Y.
Chen, Phys. Rev. D {\bf 88}, 044047 (2013).

\bibitem{Bar} J. M. Bardeen and G. T. Horowitz, Phys. Rev. D {\bf 60}, 104030
(1999).

\bibitem{Gui} M. Guica, T. Hartman, W. Song, and A. Strominger, Phys. Rev. D {\bf 80}, 124008 (2009).

\bibitem{Stro} A. Strominger and C. Vafa, Phys. Lett. B {\bf 379}, 99
(1996).

\bibitem{Hod1b} S. Hod, Phys. Rev. D {\bf 75}, 064013 (2007) [arXiv:gr-qc/0611004];
S. Hod, Class. and Quant. Grav. {\bf 24}, 4235 (2007)
[arXiv:0705.2306].

\bibitem{Gruzb} A. Gruzinov, arXiv:gr-qc/0705.1725.

\bibitem{Pescib} A. Pesci, Class. Quantum Grav. {\bf 24}, 6219 (2007).

\bibitem{Hodbb1} S. Hod, Phys. Lett. B {\bf 666} 483 (2008) [arXiv:0810.5419];
S. Hod, Phys. Rev. D {\bf 80}, 064004 (2009) [arXiv:0909.0314].

\bibitem{Hodbb2} S. Hod, Phys. Rev. D {\bf 78}, 084035 (2008)
[arXiv:0811.3806].

\bibitem{Nollert1} H. P. Nollert, Class. Quantum Grav. {\bf 16}, R159 (1999).

\bibitem{Ber1} E. Berti, V. Cardoso and A. O. Starinets, Class. Quant. Grav. {\bf 26}, 163001
(2009).

\bibitem{Kono} R. A. Konoplya and A. Zhidenko, Rev. Mod. Phys. {\bf 83}, 793 (2011).

\bibitem{Notetail} It is worth noting that the characteristic quasinormal oscillations are
usually followed by inverse power-law
decaying tails. These late-time tails represent the asymptotic
properties of the spacetime \cite{Tails1}.

\bibitem{Tails1} R. H. Price, Phys. Rev. D {\bf 5}, 2419 (1972);
E. W. Leaver, Phys. Rev. D {\bf 34}, 384 (1986); E. S. C. Ching, P.
T. Leung, W. M. Suen, and K. Young, Phys. Rev. Lett. {\bf 74}, 2414
(1995); S. Hod, Phys. Rev. Lett. {\bf 84}, 10 (2000)
[arXiv:gr-qc/9907096]; M. Tiglio, L. E. Kidder, and S. A. Teukolsky,
Class. Quant. Grav. {\bf 25}, 105022 (2008).

\bibitem{Detw} S. L. Detweiler, in Sources of Gravitational Radiation,
edited by L. Smarr (Cambridge University Press, Cambridge, England,
1979).

\bibitem{HodPRL} S. Hod, Phys. Rev. Lett. {\bf 81}, 4293 (1998)
[arXiv:gr-qc/9812002].

\bibitem{Gary} G. T. Horowitz and V. E. Hubeny, Phys. Rev. D {\bf 62}, 024027
(2000).

\bibitem{Hodqg} S. Hod, Phys. Rev. D {\bf 67}, 081501 (2003) [arXiv:gr-qc/0301122];
S. Hod and U. Keshet, Class. Quant. Grav. {\bf 22}, L71 (2005)
[arXiv:gr-qc/0505112]; S. Hod, Class. Quant. Grav. {\bf 23}, L23
(2006) [arXiv:gr-qc/0511047].

\bibitem{KeshHod} U. Keshet and S. Hod, Phys. Rev. D {\bf 76}, 061501(R) (2007) [arXiv:0705.1179].

\bibitem{KeshNe} U. Keshet and A. Neitzke, Phys. Rev. D {\bf 78}, 044006 (2008).

\bibitem{Hart} T. Hartman, W. Song, and A. Strominger, JHEP 1003:118
(2010); M. Cvetic and F. Larsen, JHEP 0909:088 (2009).

\bibitem{Noll2} H. P. Nollert, Phys. Rev. D {\bf 47}, 5253 (1993).

\bibitem{Mash} B. Mashhoon, Phys. Rev. D {\bf 31}, 290 (1985).

\bibitem{Goeb} C. J. Goebel, Astrophys. J. {\bf 172}, L95 (1972).

\bibitem{CarMir} V. Cardoso, A. S. Miranda, E. Berti, H. Witek, and V. T. Zanchin, Phys. Rev. D {\bf 79}, 064016 (2009).

\bibitem{Dolan} S. R. Dolan and A. C. Ottewill, Classical Quantum Gravity {\bf 26}, 225003 (2009);
S. R. Dolan, Phys. Rev. D {\bf 82}, 104003 (2010).

\bibitem{Massfr}  Y. D\'ecanini, A. Folacci, and B. Raffaelli, Phys. Rev. D {\bf 84}, 084035
(2011).

\bibitem{Yang} H. Yang, D. A. Nichols, F. Zhang, A. Zimmerman, Z. Zhang, and Y.
Chen, Phys. Rev. D {\bf 86}, 104006 (2012); H. Yang, F. Zhang, A.
Zimmerman, D. A. Nichols, E. Berti, and Y. Chen, Phys. Rev. D {\bf
87}, 041502(R) (2013).

\bibitem{Hodpl} S. Hod, Phys. Lett. B {\bf 715}, 348 (2012).

\bibitem{Noteunit} We use natural units in which
$G=c=\hbar=1$.

\bibitem{Notedel} Here we have used the relation $\delta\simeq
i(l+{1/2})$ for $M\omega\to 0$ \cite{Hodbb2}.

\bibitem{TeuPre1} W. H. Press and S. A. Teukolsky, Astrophys.
J. {\bf 185}, 649 (1973).

\bibitem{TeuPre2} S. A. Teukolsky and W. H. Press, Astrophys. J. {\bf 193}, 443
(1974).

\bibitem{Star} A. A. Starobinsky, Zh. Eksp. Teor. Fiz. {\bf 64}, 48 (1973) [Sov.
Phys. JETP {\bf 37}, 28 (1973)]; A. A. Starobinsky and S. M.
Churilov, Zh. Eksp. Teor. Fiz. {\bf 65}, 3 (1973) [Sov. Phys. JETP
{\bf 38}, 1 (1973)].

\bibitem{Det} S. Detweiler, Astrophys. J. {\bf 239}, 292 (1980).

\bibitem{Leaver} E. W. Leaver, Proc. R. Soc. A {\bf 402}, 285 (1985).

\bibitem{Ono1} H. Onozawa, Phys. Rev. D {\bf 55}, 3593 (1997).

\bibitem{Ander} K. Glampedakis and N. Andersson, Class. Quant. Grav. {\bf 20}, 3441 (2003).

\bibitem{Ono2} E. Berti, V. Cardoso, K. D. Kokkotas, H. Onozawa, Phys. Rev. D {\bf 68},
124018 (2003).

\bibitem{Notespin} The parameter $s$ is known as the spin weight of the
field. It is given by $s=0$ for scalar perturbations, $s=\pm {1\over
2}$ for massless neutrino perturbations, $s=\pm 1$ for
electromagnetic perturbations, and $s=\pm 2$ for gravitational
perturbations \cite{Teu}.

\bibitem{Teu} S. A. Teukolsky, Phys. Rev. Lett. {\bf 29}, 1114
(1972); S. A. Teukolsky, Astrophys. J. {\bf 185}, 635 (1973).

\bibitem{Heun} A. Ronveaux, {\it Heun's differential equations}.
(Oxford University Press, Oxford, UK, 1995).

\bibitem{Flam} C. Flammer, {\it Spheroidal Wave Functions} (Stanford
University Press, Stanford, 1957).

\bibitem{Fiz1} P. P. Fiziev, e-print arXiv:0902.1277; R. S. Borissov and P. P. Fiziev, e-print arXiv:0903.3617;
P. P. Fiziev, Phys. Rev. D {\bf 80}, 124001 (2009); P. P. Fiziev,
Class. Quant. Grav. {\bf 27}, 135001 (2010).

\bibitem{BerAlm} E. Berti, V. Cardoso, and Marc Casals, Phys. Rev. D {\bf 73}, 024013 (2006);
Erratum-ibid. D {\bf 73}, 109902 (2006).

\bibitem{Chan} S. Chandrasekhar, {\it The mathematical theory of
black holes} (Oxford University Press, Oxford, UK, 1985).

\bibitem{Page} D. N. Page, Phys. Rev. D {\bf 13}, 198 (1976).

\bibitem{Hodcen} S. Hod, Phys. Rev. Lett. {\bf 100}, 121101 (2008).

\bibitem{Abram} M. Abramowitz and I. A. Stegun, {\it Handbook of
Mathematical Functions} (Dover Publications, New York, 1970).

\bibitem{Noterel} It is worth noting that the family of polar resonances, Eq. (\ref{Eq21}),
describes perturbations of the black-hole spacetime which, in the
maximally-spinning limit $a\to M$ ($T_{\text{BH}}\to 0$), are
characterized by extremely long relaxation times:
$\tau=O(T^{-1}_{\text{BH}})\gg M$.

\bibitem{Noteqt} Specifically, one finds from Eqs. (\ref{Eq21}) and
(\ref{Eq22}) $\omega^{\text{QNM}}_n-\omega^{\text{TRM}}_{n'}=-i4\pi
T_{\text{BH}}(\ell-l)$. Using the relation $\ell=l+O(a^2\omega^2)$
[see Eq. (\ref{Eq9})] with $\omega=O(T_{\text{BH}})$ [see Eq.
(\ref{Eq21})], one finds
$\omega^{\text{QNM}}_n-\omega^{\text{TRM}}_{n'}=O(T^3_{\text{BH}})$.

\bibitem{NoteYang3} As explained above, this previously unknown family of
non-oscillatory polar QNMs was identified only recently in the
numerical analysis of Yang et. al. \cite{Yang3}.

\end{thebibliography}
\end{document}